\documentclass[aps,prc,twocolumn,floatfix,showpacs,a4paper,nofootinbib,amsmath,amssymb]{revtex4-1}
\usepackage{graphicx}
\usepackage{color}
\usepackage{dcolumn}
\usepackage{bm}
\newcommand{\be}{\begin{equation}}
\newcommand{\ee}{\end{equation}}
\newcommand{\ba}{\begin{eqnarray}}
\newcommand{\ea}{\end{eqnarray}}
\newcommand{\bd}{\begin{displaymath}}
\newcommand{\ed}{\end{displaymath}}
\renewcommand{\vec}[1]{\mbox{\boldmath$#1$}}

\begin{document}

\title{Rotation in an exact hydro model}
\author{L.P. Csernai, D.J. Wang}
\affiliation{ Institute of Physics and Technology, University of Bergen,
Allegaten 55, 5007 Bergen, Norway }
\author{T. Cs\"org\H o}
\affiliation{
Wigner Research Center for Physics,
H-1121 Budapesti 114, POBox 49, Hungary}
\date{\today}
\begin{abstract}
We study an exact and extended solution of the fluid dynamical
model of heavy ion reactions, and  estimate the rate of slowing
down of the rotation due to the longitudinal and transverse
expansion of the system. The initial state parameters of the
model are set on the basis of a realistic 3+1D fluid dynamical
calculation at TeV energies, where the rotation
is enhanced by the build up of the Kelvin Helmholtz Instability
in the flow.
\end{abstract}


\maketitle

\section{Introduction}

In peripheral heavy ion collisions the system has angular momentum.
It has been shown in hydrodynamical computations that
this leads to a large shear and vorticity \cite{CMW13}.
Furthermore when the Quark-Gluon Plasma
(QGP) is formed with low viscosity \cite{CKM},
interesting new phenomena may occur like rotation
\cite{hydro1}, or turbulence,
which shows up in form of a starting Kelvin-Helmholtz
Instability (KHI) \cite{hydro2,WNC13}.

Surprisingly, the effects arising from non-vanishing
initial angular momentum in fireball hydrodynamics can be studied
with the help of exact and explicit, analytic solutions of
the equations of hydrodynamics. The pioneers to the application
of the hydrodynamical method to high energy
particle and nuclear collisions
initially neglected the effects from the
non-vanishing angular momentum:
Belenkij and Landau considered the 1+1 dimensional
explosion of a non-expanding
region of very high energy
density~\cite{Belenkij:1956cd},
while Hwa and Bjorken
considered ~\cite{Hwa:1974gn,Bjorken:1982qr}
a boost-invariant, already asymptotic expansion
as the initial condition of 1+1 dimensional
expansion governed by relativistic hydrodynamics.

However, if the collision of two protons
or two heavy ions is not exactly head-on,
there is a non-vanishing initial angular
momentum present in the initial conditions,
which was for a long time neglected. However,
rather recently, numerical investigations
of relativistic
hydrodynamics~\cite{CMW13,CKM,hydro1,hydro2,WNC13},
as well as exact and explicit analytic solutions of
relativistic and non-relativistic
hydrodynamics were found for rotating fluids
~\cite{Nagy:2009eq,Csorgo:2013ksa}.
It took 35 years after the publication of the first
exact, non-rotational solution
in a similar class~\cite{Bondorf:1978kz}
to find and publish rotating solutions of hydrodynamics,
a long road with lots of surprises and unexpected turns,
that were recently briefly summarized
in ref. ~\cite{Csorgo:2013ksa}.
In ref. \cite{McI-Teo} the AdS/CFT holography
method is used to study the QGP created
in heavy ion collisions. The authors chose two types of velocity 
profiles, which do not favor instability classically, and they use
the planar black hole geometry to compute the fluid shear by using 
parameters predicted in \cite{hydro2}. 
The discussion shows that for the two clasically stable configurations
in the holographic model instability 
develops very slowly for lower chemical potentials, 
but turbulence may still exist for high chemical potentials.
Finding of these rotating solutions started to
shed more light on the effects of rotation,
which plays a very important role
\cite{McI2014}
in shaping the events in astrophysical, hydrodynamically
evolving systems, turbulence, or in eddy and vortex formation
in fluid dynamics.
This motivates the subject of the present
paper, which evaluates numerically
some of the important characteristics of the
fireball (like the size of the fireball
along the axis of rotation and the size of
the fireball in the plane perpendicular
to the rotation) for quasi-realistic initial
conditions and compares it with the
case when the initial angular momentum is
negligible, or is neglected.

Based on ref. \cite{hydro2} we can extract some basic parameters
of the rotation obtained with numerical fluid dynamical model
PICR.  These parameters are extracted from model calculation
of a Pb+Pb collisions
at $\sqrt{s_{NN}} = 2.76$ A TeV and at the impact parameter of
$b=0.7 b_{max}$, with high resolution and
thus small numerical viscosity. Thus, in this collision
the KHI occurs and enhances rotation at intermediate times,
because the turbulent rotation gains energy from the
original shear flow. The turbulent rotations leads to a
rotation profile where the rotation of the external regions
lags behind the rotation of the internal zones. This is
a typical growth of the KHI. See Table \ref{t1}.

\bigskip

\begin{table}[h]   
\begin{tabular}{cccccccc} \hline\hline \phantom{\Large $^|_|$}
$t$  &  $Y$ &$\dot{Y}$ &  $\theta$ &$R$ &$\dot{R}$& $\omega$ \\
(fm/c)& (fm)& (c)   &  (Rad)  &  (fm)  &  (c)    & (c/fm) \\
\hline
0.   & 4.38 & 0.90  &  0.000  &  3.68  &  0.60   &  0.0175 \\
2.   & 6.18 & 0.88  &  0.035  &  4.87  &  0.84   & 0.0350 \\
4.   & 7.91 & 0.84  &  0.105  &  6.56  &  0.97   & 0.0520 \\
6.   & 9.54 & 0.80  &  0.209  &  8.49  &  0.86   & 0.0700 \\
8.   &11.09 & 0.76  &  0.349  & 10.21  &  0.81   & 0.0350 \\
\hline
\end{tabular}
\caption{
Time dependence of some characteristic parameters of the
fluid dynamical calculation presented in ref. \cite{hydro2}.
$R$ is the average transverse radius ($R \approx \sqrt{X Z}$),
$Y$ is the length of the system in the direction of the
axis of the rotation $y$, $\theta$ is the polar angle of the rotation
of the interior region of the system measured versus the 
 $z$ directed, beam axis, of the reaction plane, $[x,z]$ plane,
$\dot{R},\ \dot{Y}$ are the speeds of expansion in radial and
rotational axis directions, and $\omega$ is the angular velocity of the
internal region of the matter during the collision.
}
\label{t1}
\end{table}   

The initial angular momentum of the system is large,
$ L_y = - 1.05\times 10^4 \hbar$ in mid-peripheral collision. 
This is arising from the pre-collision state. In refs. 
\cite{CMW13,hydro1,hydro2} 
the same Initial State (IS) model is used,
before the PICR fluid dynamical model started. The IS model
\cite{M1,M2} 
describes the first 4 fm/c time period after the moment when
the Lorentz contracted, thin projectile and target slabs
(like in the CGC model)  penetrate each other. 
The matter is then divided into longitudinally ($z$ directed)
expanding {\it (fire)streaks}. The $[x, y]$ transverse plane is divided 
into surface elements according to the resolution of the PICR model,
and each element belongs to one streak. The streak expansion is
described in a one-dimensional classical Yang-Mills field, spanned
by the color charges at the two ends of the streak. This field 
slows down the expansion due to the large {\it string-rope tension}.
The IS dynamics satisfies the  momentum conservation $Streak\ by\ Streak$, 
and this way the total angular momentum of the IS is 
also exactly conserved. During the IS model the dynamics is
one dimensional ($z$), with $z$ directed velocities only, but the
expansion is different in different transverse points.

After 4 fm/c of the IS model, local equilibrium is reached and the
PICR (3+1)D fluid dynamical model starts (with $t=0$ fm/c fluid dynamical
model time). Then, due to the fluid dynamical development and equilibration, 
the $x$-directed velocity starts to increase and the average of the
$z$-directed velocity decreases. This way the angular momentum is exactly
conserved in the (3+1)D fluid dynamical calculation.
Thus the $x$-directed velocity or otherwise the rotation in the 
horizontal plane starts up delayed. This is a fully realistic model 
of the initial longitudinally transparent non-equilibrium 
dynamics, and the subsequent equilibration of rotation on a larger scale
\cite{hydro2}.

As the initial conditions and initial times in the exact model 
and in the PICR (3+1)D model are not identical
we have matched the time coordinates such that $t_{exact}=0$
 fm/c in the exact model corresponds to: $t = 3$ fm/c in
the full (3+1)D calculation.

If we compare the rotation of the horizontal plane 
($x$ directed velocity) only, then the PICR model and 
the exact model are becoming similar 
at $t=6$ ($t_{exact}=3$) fm/c and after (see Table I and Table II). 
If however, we estimate the average of the $z$ directed velocity 
also, and consider then the average of the $x$ and $z$ directed 
velocities, the two models are becoming similar already
at $t=4$ ($t_{exact}=1$) fm/c and after.  Thus, the applicability
of the exact model with the parameters chosen here, starts 
approximately after $t=5$ fm/c on the timescale of the PICR model.
The radius, $R$, parameters are matched to each other in the two models s0
that in the (3+1)D model at the same time moments, $t=5$ and $8$ fm/c the
radii are 8.49 and 10.21 fm for a sharp matter surface, while in the exact 
model the corresponding radii are 3.97 and 5.36 fm respectively 
(i.e. about half of the previous values) but these represent the width
parameter of an infinite Gaussian matter distribution. 

The initial part of the (3+1)D model describes the equilibration of the
rotational flow from the initial shear flow, which rotation then
leads to a maximal, azimuthally averaged 
angular velocity. Then the system expands and the angular velocity 
decreases. The exact model, assuming uniform rotation can only describe
this second phase of the process. It is important to mention that
the KHI facilitates the equilibration and speed up of the rotation and leads
to an earlier and bigger maximal angular velocity. In this case the
applicability of the exact model is more extended in time, and spans
the range between the equilibration of the rotation and the freeze out.
At lower beam energies and small impact parameters (i.e. at lover 
angular momentum) the timespan of the applicability of exact model
should be tested separately.

We want to use these fluid dynamical calculations to test a new family
of exact rotating solutions \cite{Csorgo:2013ksa},
which may provide more fundamental
insight to the interaction between the rotation and
expansion of the system.
This model offers a few possible variations,
here we chose the version {\em 1A} to test.
We change the axis labeling of ref. \cite{Csorgo:2013ksa}, so that
the axis of the rotation is $y$ while the transverse plane of the
rotation is the $[x,z]$ plane. Thus the values extracted from the
results of the fluid dynamical model, \cite{hydro2}, should take
this into account. The initial radius parameter, $R$, corresponds
to the system size in the $x$ or $z$ directions in hydro,
and we assume an $x,z$ symmetry in the
exact model. The rotation axis is the $y$ axis in hydro and now also
in the exact model. The exact model assumes azimuthal
symmetry, so it cannot describe the beam directed elongation of
the system, but this is arising from the initial beam momentum,
and we intend to describe the rotation of the interior part
of the reaction plane and the rotation there.

In Section \ref{S2} we recapitulate some of the central results
of ref. \cite{Csorgo:2013ksa} for clarity and so that the
manuscript be self-contained. New studies will start in Section \ref{S3}.

\section{From the Euler Equation to Scaling}
\label{S2}

In ref. \cite{Csorgo:2013ksa} it is assumed that
the temperature and the density have time independent distributions
with respect to a scaling variable:
$
 s = r_x^2/X^2 + r_y^2/Y^2 + r_z^2/Z^2 \ .
$
Now instead we assume azimuthal symmetry and thus
use the corresponding
cylindrical coordinates instead of $(x, y, z)$. However,
we also use the coordinates in length
dimension, $(r_\rho, r_\varphi, r_y)$,
so that
$$
r_\rho = \rho,\ \  r_\varphi= r_\rho \varphi,\ \  r_y = y \ .
$$
These represent the so called "out, side, long" directions.
The boundary values of these coordinates are then $(R, S, Y)$.
The scaling variables can be introduced as
$$
s_\rho    =   r_\rho^2/R^2, \ \
s_\varphi =   r_\varphi^2/S^2 \ \
s_y       =   r_y^2/Y^2 \ ,
$$
where $S$ is the roll-length on the outside circumference,
 starting from $\varphi_{0} = 0$ and $S_0=0$ at $t_0$, \ \
$S = R \varphi$  and $\dot{\varphi}~=~\omega$ and this
displacement is orthogonal to the longitudinal and transverse
displacements. The internal roll-length is
$r_\varphi   = \varphi\,   r_\rho $, the corresponding velocity is
$v_\varphi   = \omega\,    r_\rho $, and so
$v_\varphi^2 = \omega^2\,  r_\rho^2 $. On the other hand from the
scaling of $r_\rho$, it follows that $ r_\rho^2 = R^2\, s_\rho$.

  Nevertheless, in case of these
scaling variables the distributions of density and temperature,
$n(s)$ and $T(t,s)$
should not depend on $s_\varphi$ or  $r_\varphi$, just on the radius and the
longitudinal coordinates. Therefore in this work we introduce another
scaling variable:
$$
s \equiv s_\rho + s_y\ .
$$

Our reference frame is then spanned by the
directions $(r_\rho, r_\varphi, r_y)$.
In this case due to the azimuthal symmetry the derivatives,
$\partial s/\partial r_\varphi$ vanish.
In this coordinate system the volume is $V = 2 \pi  R^2 Y$.

The derivatives,
$\dot{R}(t_0)$ and $\dot{Y}(t_0)$ in this exact model should not
equal the ones obtained from the fluid dynamical model,
because in the
more realistic model the density and velocity profiles do not agree with
the exact model's assumptions.
Also initially in the realistic
fluid dynamical model the angular momentum increases in the
central region due to the developing turbulence, while in the exact model
it is monotonously decreasing due to the scaling expansion.

For simplicity we also
assume that the Equation of State (EoS), $\epsilon =  \epsilon(n,p)$,
 with a constant $\kappa$ is:
\be
\epsilon = \kappa p \ \ \ {\rm and} \ \ \ p = nT \ ,
\ee
\noindent
where $n$ is the conserved net baryon charge and $T$ is the temperature.

Now we calculate the eq. (15) in
ref. \cite{Csorgo:2013ksa}.
\be
n\,m\, (\partial_t + \vec{v}\cdot\nabla) \vec{v} = - \nabla p \ .
\label{Euler}
\ee
For the variables of this equation we have:
\ba
T &=&T_0\left(\frac{V_0}{V}\right)^{1/\kappa}\mathcal{T}(s)\ ,
\nonumber \\
n &=&n_0\frac{V_0}{V}\nu (s) ,
\nonumber \\
\nu(s) &=& \frac{1}{\mathcal{T}(s)}
e^{-\frac{1}{2}\int_0^s\frac{d u}{\mathcal{T}(u)}},
\label{defs}
\ea
and in addition
in ref. \cite{Csorgo:2013ksa} it is assumed that
the temperature and the density have time independent distributions
with respect to the scaling variable $s$.


Thus, for the {\bf right hand side} of eq. (\ref{Euler}):
\ba
-\nabla p &=& -\nabla n T \nonumber\\
&=& - n_0\frac{V_0}{V}T_0\left(\frac{V_0}{V}\right)^{1/\kappa}
 \nabla e^{-\frac{1}{2}\int_0^s\frac{du}{\mathcal{T}(u)}}
\nonumber\\
&=& - n_0\frac{V_0}{V}T_0\left(\frac{V_0}{V}\right)^{1/\kappa}
     e^{-\frac{1}{2}\int_0^s\frac{du}{\mathcal{T}(u)}}
     (-\frac{1}{2})\frac{1}{\mathcal{T}(s)}\nabla s
\nonumber\\
&=&n m Q / V^\gamma
\left(\frac{r_\rho}{R^2}\vec{e}_\rho {+}\frac{r_y}{Y^2}\vec{e}_y \right) \ ,
\label{eul}
\ea
where $\gamma = 1/\kappa$ and $Q\equiv\frac{T_0 V_0^\gamma}{m}$.


Using the $\rho, \varphi, y$ coordinates, the rotation
would show up as an independent orthogonal term.
However,
the closed system has no external torque, and the
internal force from the gradient of the pressure is
radial, which does not contribute to tangential acceleration.
The change of the angular velocity arises from the angular momentum
conservation in the closed system as a constraint, so we do not have
to derive additional dynamical equations to describe the evolution
of the rotation.

Now for the {\bf left hand side} of eq. (\ref{Euler}),
the velocity, $\vec{v} =\vec{v}(t,r_\rho, r_\varphi, r_y )$, scales as
\be
\vec{v} =
v_\rho \vec{e}_\rho {-} v_\varphi \vec{e}_\varphi {+} v_y \vec{e}_y =
\frac{\dot{R}}{R} r_\rho \vec{e}_\rho {-}
\omega r_\rho \vec{e}_\varphi {+}
\frac{\dot{Y}}{Y} r_y \vec{e}_y \ .
\ee
Let us first calculate the time derivatives for the components.
(See e.g. \cite{Stoecker_handbook}):
\ba
\partial_t v_\rho &=& \left[ \left(
\frac{\ddot{R}}{R}{-}\frac{\dot{R}^2}{R^2}\right) - \omega^2 \right]  r_\rho ,
\nonumber \\
\partial_t v_\varphi &=&
- \omega \frac{\dot{R}}{R} r_\rho , \ \ \
\partial_t v_y =
\left[ \frac{\ddot{Y}}{Y}{-}\frac{\dot{Y}^2}{Y^2}\right]  r_y .
\label{partialtvx}
\ea
The other term of the comoving derivative gives:
\be
(\vec{v}\cdot\nabla) \vec{v} =
\frac{\dot{R}^2}{R^2}r_\rho     \vec{e}_\rho +
\omega \frac{\dot{R}}{R} r_\rho \vec{e}_\varphi +
\frac{\dot{Y}^2}{Y^2}r_z \vec{e}_z \ .
\label{vDeltvx}
\ee
Adding eq. (\ref{partialtvx}) and (\ref{vDeltvx}) we get:
\ba
mn(\partial_t{+}\vec{v}\cdot\nabla)v_\rho
 &=& mn\left[\left( \ddot{R}/R\right) - \omega^2 \right] r_\rho \ ,
\nonumber \\
mn(\partial_t{+}\vec{v}\cdot\nabla)v_z
 &=& mn\left( \ddot{Y}/Y\right) r_z \ .
\ea
Then the {\bf equality} of the right hand side and left hand side
of the Euler equation (\ref{Euler}) leads
to the ordinary differential equations.
Multiplying the two non-vanishing equations with $R^2$ and $Y^2$
respectively yields:
%

%
\ba
R\ddot{R} - W/R^2 &=& Y\ddot{Y} =
 \frac{T_0}{m}\left(\frac{V_0}{V}\right)^{\gamma} \ ,
\label{e-2s}
\ea
where $W~\equiv~\omega_0^2~R_0^4$.
Notice that from the angular momentum conservation
$\omega=\omega_0R_0^2/R^2$, thus the rotational
term, $R^2\omega^2$ in the equation,
takes the form  $W / R^2$.

Notice that due to the EoS the pressure is proportional to the
baryon density $n$, just as the r.h.s. of the Euler equation,
therefore the equation of motion does not depend on $n$ or $n_0$.

\section{Conservation Laws}
\label{S3}

If we want to calculate the energy of the whole system,
then we should actually integrate it for the whole volume, $V$.
Thus, not only the scaling of
$\vec{v} = ( v_\rho, v_\varphi, v_z)$ but the particle density
distribution, $n(s)$ will also play a role.
\smallskip

The rotational energy at the {\bf surface} is
$\mathcal{E}_{Side} \equiv \frac{1}{2} m \dot{S}^2 =
 \frac{1}{2} m R^2 \omega^2$,
and if we express $\omega$ via $\omega_0$ using the relation
$\omega = \omega_0 R_0^2 / R^2$, then
$\mathcal{E}_{Side} = W / R^2$, i.e. just as before.
The expansion energy at the surface is
$\mathcal{E}_{Out} \equiv \frac{1}{2} m \dot{R}^2$, and for the
longitudinal direction,
$\mathcal{E}_{Long} \equiv \frac{1}{2} m \dot{Y}^2$.
\smallskip

For the interior we can calculate the radial and longitudinal
expansion velocities and the corresponding kinetic energies, as
well as the kinetic energy of the rotation.
In the evaluation of the internal and kinetic
energies the radial and longitudinal density profiles of the
system should be taken into account.

{\em We now assume that the temperature
profile is flat, and consequently the density profiles are Gaussian and
separable} \cite{ACLS01}.
With this approximation the different integrated energies are calculated.
The boundary of the spatial integrals can be set to infinity 
or to finite values ($s_{\rho M},\ s_{yM}$) as well. To be 
consistent with the earlier
exact model results we integrate now to infinity, the integrals are finite.

Adding up the kinetic energies yields
\be
E_K = \frac{1}{2} m N_B \left(
\alpha^2 \dot{R}^2 +  \alpha^2 R^2 \omega^2 + \beta^2 \dot{Y}^2 \right) \ ,
\ee
where\\
$
\alpha^2\! \equiv \!
4\sqrt{2}\,C_n I_B({\small \frac{1}{2}}s_{yM}) 
I_C({\small \frac{1}{2}}s_{\rho M})\           {\rm and}\, \\
\beta^2\! \equiv\!
4\sqrt{2}\,C_n I_A({\small \frac{1}{2}}s_{\rho M}) 
I_D({\small \frac{1}{2}}s_{yM}), {\rm where}\\
$
$C_n = 1 \left/ \left[  2\sqrt{2}\, I_A(0.5s_{\rho M})\, I_B(0.5s_{yM}) \right] \right.$\
\footnote{
$I_A(u) = 1 - \exp(-u)$,
$I_B(u) = \sqrt{\pi}\, \Phi(\sqrt{u}\,)$,
$I_C(u) = 1 - (1+u) \exp(-u)$,
$I_D(u) = \frac{\sqrt{\pi}}{2} \Phi(\sqrt{u}) - \sqrt{u} e^{-u}$,
where
$\Phi(u)= {\rm erf}(u) \equiv \frac{2}{\sqrt{\pi}} \int_0^u \exp(-x^2)\,dx$
\ \cite{IGF}.}
.
Now we extend the boundaries to infinity, thus 
$\alpha^2 = 2$ and $\beta^2 = 1$.

Here $\alpha^2$ and $\beta^2$ are time independent, because
they depend on the scaling variables only.
If we divide this result by the conserved baryon charge, $N_B$, we will get
\be
\frac{E_K}{N_B} =  \frac{1}{2} m  \left[
 \alpha^2 \left( \dot{R}^2 {+} R^2 \omega^2 \right)
 +   \beta^2\, \dot{Y}^2 \right]\ .
\ee

Based on the EoS, $\epsilon = \kappa p = \kappa n T$, one can
calculate the compression energy also based on the
density profiles of $n(s)$ and $\epsilon(s) = \kappa\, n(s) T$.
Here we made the same simplifying assumptions on the density profiles
as before.

Then volume integrated internal energy and net baryon charge
will have the same density profile, normalized to $N_B$:
\vskip -2mm
\ba
E_{int} &=& \kappa  \int p dV = \kappa \int n T dV =
\kappa N_B T_0 (V_0/V)^\gamma \, C_n
\nonumber \\
 &\times& \frac{1}{V}\ 2 \pi R^2Y\!\! \int_0^{s_{yM}} \!\! 
\int_{0}^{s_{\rho M}}\!\!\!
\nu(s)\ ds_\rho\, \frac{ds_y}{\sqrt{s_y}}
\nonumber \\
 &=&  \kappa  N_B T_0 (V_0/V)^\gamma  \ = \
 \kappa T_0\left(\frac{V_0}{V}\right)^{\gamma}
\ea
where $C_n$ is the normalization constant.

\section{Reduction to a Single Differential Equation}

Now following the method of ref. \cite{ACLS01} ,
we  study the following combination:
\ba
{\cal F} &=& {\small \frac{1}{2}} \partial_t^2
\left( \alpha^2 R^2 +  \beta^2 Y^2 \right)
 = \partial_t \left(
 \alpha^2 R \dot{R}  + \beta^2 Y \dot{Y} \right)
\nonumber\\
 &=& \alpha^2 \dot{R}^2  + \beta^2 \dot{Y}^2 +
\alpha^2 R \ddot{R}  + \beta^2 Y \ddot{Y}  .
\label{F16}
\ea

Here we used the notation
$
\partial_t = \frac{\partial}{\partial t} \ \ {\rm and } \ \
\partial_t^2 = \frac{\partial^2}{\partial t^2}.
$
Now we can replace the last two terms,
$ \alpha^2 R \ddot{R},\  \beta^2 Y \ddot{Y}$, by
using eqs. (\ref{e-2s}), i.e. we use the Euler equation (\ref{Euler}). Then
we obtain:
\be
 {\cal F} =
  \alpha^2 \dot{R}^2 + \beta^2 \dot{Y}^2 + \alpha^2 \frac{W}{R^2}
+ (\alpha^2{+}\beta^2) \frac{Q}{(2 \pi R^2Y)^\gamma} ,
\label{EC1}
\ee
On the other hand from the {\bf energy conservation},
$E_{tot} = E_k + E_{int}$, we get that
\be
\frac{E_{tot} }{N_B \, m} = \frac{1}{2}
\left[
 \alpha^2 \dot{R}^2 {+} \beta^2 \dot{Y}^2 {+} \alpha^2 \frac{W}{R^2}
{+} \frac{2 \kappa Q}{(2 \pi R^2Y)^\gamma}
\right] ,
\label{EC2}
\ee
where we used the EoS and thus the parameter $\kappa$ now appears
in the expression of the energy.

Now, if our EoS is such that
\be
     \kappa = (\alpha^2{+}\beta^2)\ / 2\ ,
\ee
then $ {\cal F} = 2 E_{tot} / (N_B \, m ) = $ const., and in the same
type of calculation
as in ref. \cite{ACLS01}, we can introduce
\be
U^2(t) \equiv  \alpha^2 R^2(t) +  \beta^2 Y^2(t) ,
\label{eu}
\ee
which satisfies
\be
 \partial_t^2 \left( \alpha^2 R^2 + \beta^2 Y^2 \right) =
 \partial_t^2 U^2(t) = 2 {\cal F}  \ .
\label{u21}
\ee
Thus, the solution of eq. (\ref{u21}), can be parametrized as:
\be
U^2(t) = A (t-t_0)^2 + B (t-t_0) + C \ ,
\ee
\ba
{\rm where:}\ \
A&=& \alpha^2 \dot{R}_0^2+\beta^2\dot{Y}_0^2+ \alpha^2 W/ R_0^2 +
(\alpha^2{+}\beta^2) \frac{T_0}{m}
\nonumber \\
B&=& 2\alpha^2 R_0 \dot{R}_0 + 2\beta^2 Y_0 \dot{Y}_0
\phantom{\frac{T_0}{m}}
\nonumber \\
C&=&  \alpha^2 R_0^2 +  \beta^2 Y_0^2     \phantom{\frac{T_0}{m}}\ .
\ea
\begin{table}[h]       
\begin{tabular}{cccccccc} \hline\hline \phantom{\Large $^|_|$}
$t$  &  $Y$ & $\dot{Y}$ & $\theta$ & $R$ & $\dot{R}$ & $\omega$ \\
(fm/c)&\ \ (fm)\ \ &\ \ (c)\ \ \ & \ (Rad)\ \ & \ \ (fm)\ \ &\ \ (c)\ \ \ &\ (c/fm)\ \\
\hline
0.0 & 4.000 & 0.300 & 0.000 & 2.500 & 0.250 & 0.150 \\
1.0 & 4.349 & 0.393 & 0.135 & 2.852 & 0.441 & 0.115 \\
2.0 & 4.776 & 0.458 & 0.235 & 3.360 & 0.567 & 0.083 \\
3.0 & 5.258 & 0.503 & 0.307 & 3.970 & 0.646 & 0.059 \\
4.0 & 5.777 & 0.534 & 0.358 & 4.642 & 0.696 & 0.044 \\
5.0 & 6.322 & 0.555 & 0.397 & 5.356 & 0.729 & 0.033 \\
6.0 & 6.886 & 0.571 & 0.426 & 6.096 & 0.752 & 0.025 \\
7.0 & 7.462 & 0.582 & 0.449 & 6.856 & 0.767 & 0.020 \\
8.0 & 8.049 & 0.591 & 0.467 & 7.629 & 0.779 & 0.016 \\
\hline
\end{tabular}
\caption{
Time dependence of characteristic parameters of the exact
fluid dynamical model \cite{Csorgo:2013ksa}.
R is the transverse radius, Y is the
(rotation axis directed) length  of the
system, $\dot{R},\ \dot{Y}$ are the speed of expansion in transverse and
axis directions, $\theta$ id the angle of rotation, and $\omega$ is the 
angular velocity of the matter.
}
\label{t2}
\end{table}           
Let us take one of the Euler equations from eq. (\ref{e-2s}),
\be
\ddot{Y} =  \frac{Q}{Y (2 \pi R^2 Y)^{\gamma}} \ ,
\label{e-4s}
\ee
and express $R^2$ in terms of $U^2(t)$ which is known
based on the energy conservation:
\be
R^2(t) = (U^2(t) - \beta^2 Y^2)/ \alpha^2 \ .
\label{er}
\ee

\begin{figure}[h]     
\begin{center}
\resizebox{0.9\columnwidth}{!}
{\includegraphics{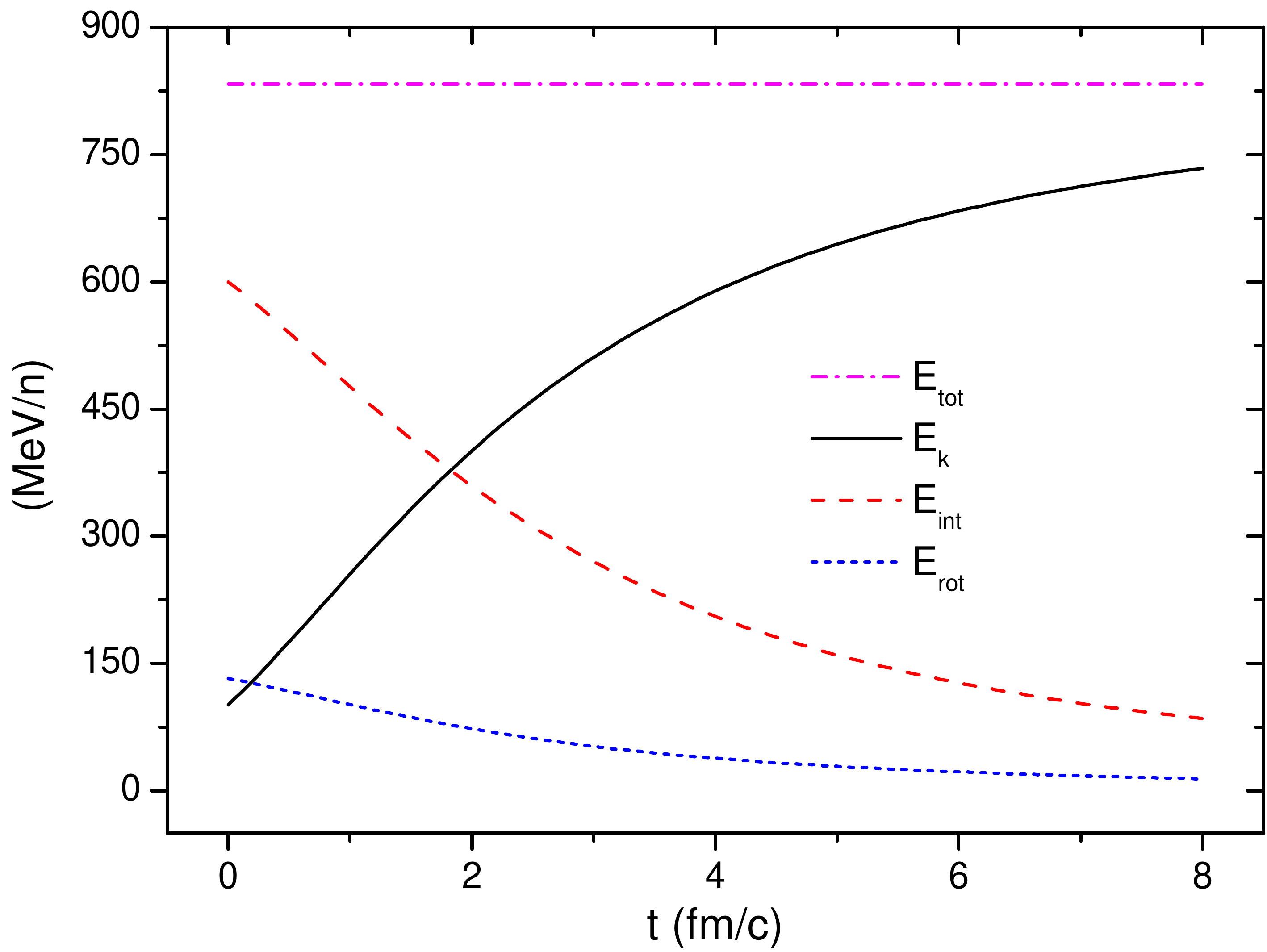}}
\caption{
(color online)
The time dependence of the kinetic energy of the expansion, $E_K$,
of the internal energy, $E_{int}$, and the rotational energy,
$E_{rot}$ per nucleon in the exact model with the initial conditions
described above. The kinetic energy of the expansion is increasing,
at the cost of the decreasing internal energy and the slower decreasing
rotational energy. The rotational energy is decreasing to the half
of the initial one in 2.1 fm/c.
}
\label{F1-E-vs-t}
\end{center}
\end{figure}           
\vskip -3mm

This leads to a 2nd order differential equation for $Y(t)$:
\be
\ddot{Y} =
\frac  {\alpha^{2\gamma}\, Q}
{ Y \left[2 \pi Y (U^2(t) {-} \beta^2 Y^2)\right]^{\gamma}}
= f(Y,t)\ ,
\label{e-5s}
\ee
which can be solved.
Then $R(t)$ and $\dot{R}(t)$ are  given
by eqs. (\ref{er}) and (\ref{EC1}) respectively.

In the main steps we followed ref. \cite{ACLS01}, however it turned out
that the modified last step of the method provides a more
straightforward solution. We show the model provides an excellent
and simple semi-analytic tool to study the effects and
consequences of an expanding and rotating system.

We used the Runge Kutta \cite{RungeKutta} method to solve this
differential equation. We need the constants, $Q$ and $W$, as well
as the initial conditions for $R$ and $Y$.

Based on fluid dynamical
model calculation results, presented in table \ref{t1}, we chose the
parameters:
$
T_0 = 250\  {\rm MeV},\
m  =  939.57 {\rm MeV}\
\omega_0 =0.15$ c/fm.
For the internal region we take the initial radius parameters as
$
R_0 = 2.5 {\rm fm\ and\ }
\dot{R} = 0.25 {\rm c}
$,
and we disregard the larger extension
in the beam direction, because our model is azimuthally
symmetric and because the beam directed
large elongation is a consequence
of the initial beam directed momentum excess, which is converted into
rotation in the course of initial equilibration only.
In this exact model the rotation axis, denoted by $Y$, corresponds to the
out of plane, $y$ direction in the fluid dynamical model
(and not to the beam direction!). Due to the eccentricity at finite
impact parameters, with an almond shape profile, the initial out of plane
size is larger than the in plane transverse size, so we chose
$
Y_0 = 4.0  {\rm fm\ and\ }
\dot{Y} = 0.3 {\rm c}
$.
\begin{figure}[ht]     
\begin{center}
\resizebox{0.9\columnwidth}{!}
{\includegraphics{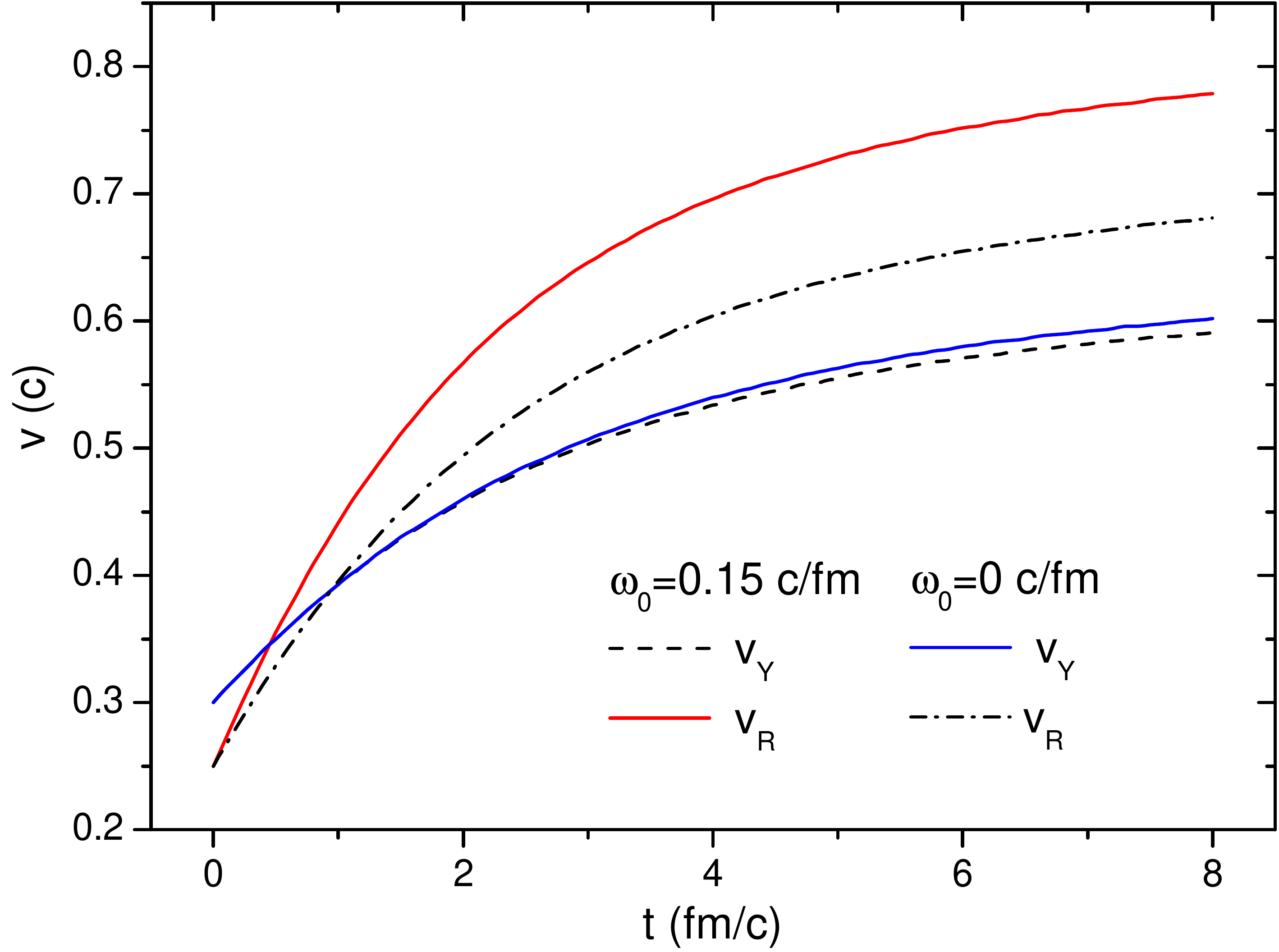}}
\caption{
(color online)
The time dependence of the velocity of expansion in the
transverse radial direction, $v_R$ and in the direction of
the axis of the rotation, $v_Y$.
}
\label{F2-v-vs-t}
\end{center}
\end{figure}        
%
As the exact solution is able to describe the monotonic expansion,
and so the steady decrease of the rotation, we start from a higher
initial angular velocity than shown by the fluid dynamical model, PICR,
as the angular velocity was measured versus the horizontal plane
where the angular velocity starts from zero.

With these initial parameters the exact model yields a dynamical development
shown in Table \ref{t2}. According to expectations the radius, $R$, and the
axis directed size, $Y$, are increasing, the angular velocity, $\omega$
decreases, The total energy is conserved, while the kinetic
energy of expansion is increasing, and that of the rotation and
internal energy are decreasing.
See Fig. \ref{F1-E-vs-t}.

The change of the expansion velocity is shown in Fig.  \ref{F2-v-vs-t}.
The expansion velocity
is increasing in both directions. While in the axis, $y$, direction
the velocity increases from 0.4 c to 0.5 c in 8 fm/c time, the
radial expansion increases faster, in part due to the centrifugal
force from the rotation.
The radial expansion velocity increases by near to 10 percent due to
the rotation, which is significant, while the expansion in the direction
of the axis of rotation is hardly changed. In both cases the expansion
in the radial direction is large. This is due to the choice of a small
initial radius parameter.
This exact, perfect fluid  model overestimates the radial expansion
velocity due to the lack of dissipation and the Freeze out happens
earlier than 8 fm/c as at this time the size of the system is already
reaching 16 fm (see Fig. 3) larger than the estimates based two
particle correlation experiments.
At the same time, although the rotation is non-relativistic, the Hubble
flow terms are large and the Hubble flows are relativistic. This problem
is not new, it is also present in ref. [15] and in earlier exact models.
The radial (directional Hubble) solutions go smoothly over to a
relativistic exact solution of hydrodynamics  \cite{CsCsHK03}.
As the velocity of rotation decreases faster asymptotically 
than the Hubble flow, the rotation is not expected to influence
essentially the asymptotic relativistic behaviour of the flow.

\begin{figure}[h]      
\begin{center}
\resizebox{0.9\columnwidth}{!}
{\includegraphics{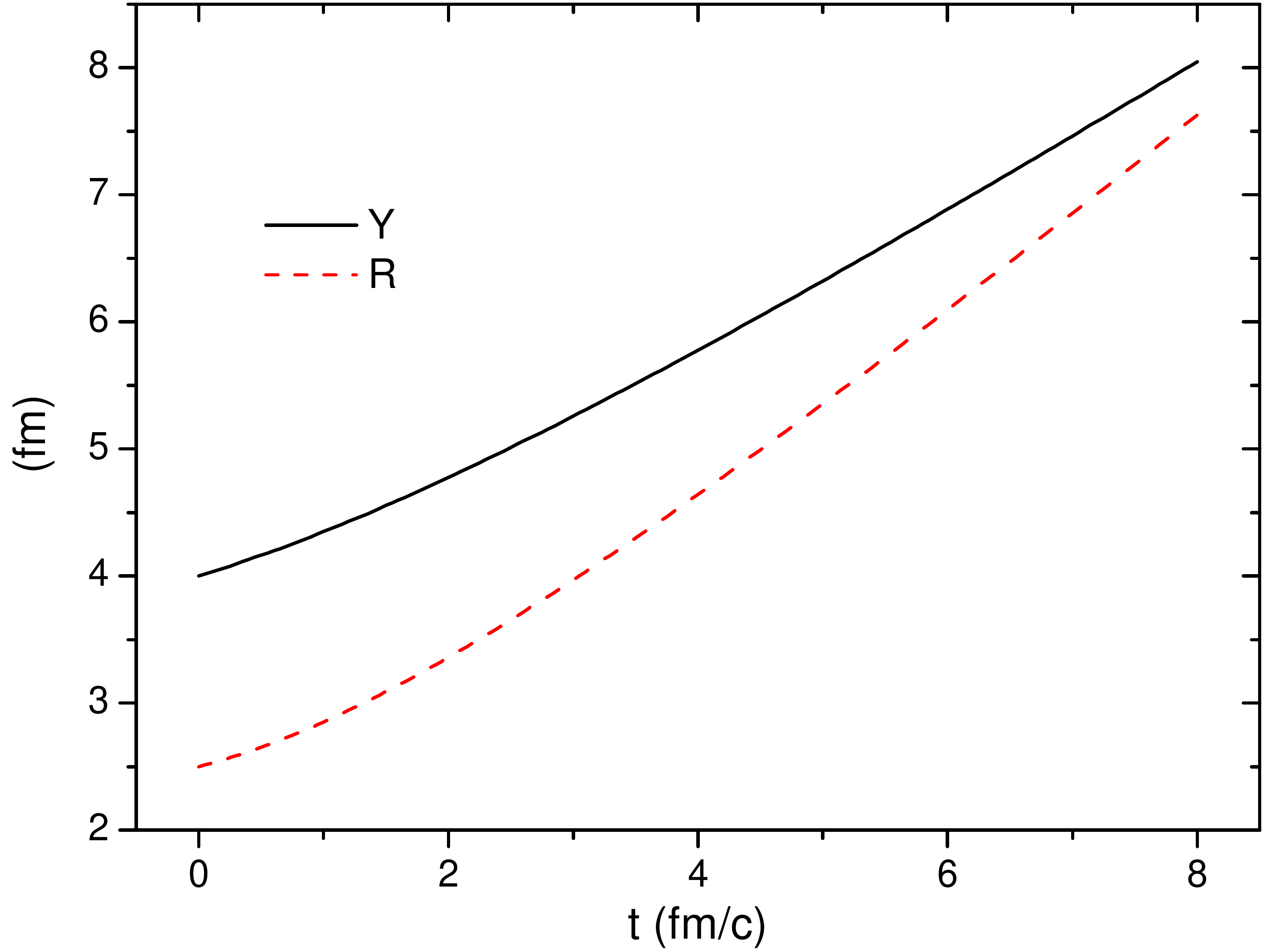}}
\caption{
(color online)
The time dependence of the Radial, $R$, and the $y$ axis directed
size, $Y$, of the expanding system.
As the $y$ directed velocity is initially larger and its change
is relatively smaller the change of the rate of increase is
hardly visible.
}
\label{F3-RY-vs-t}
\end{center}
\end{figure}

The more rapid velocity change arises partly from the centrifugal
acceleration of the rotation, but also from the fact that the
initially smaller transverse size increases faster in the direction
of equal sizes in both directions. See Fig.  \ref{F3-RY-vs-t}.
%

\section{Conclusions}
%
In conclusion, the exact model can be well realized with
parameters extracted from detailed, high resolution, 3+1D
relativistic fluid dynamical model calculations with the
PICR code. 
The exact model describes a system with  with one single, uniform $\omega$
representing the whole matter at a given moment of time. Depending on the
impact parameter, the system size, the beam energy, and the transport
properties, the uniform flow can develop from the initial shear flow
at different times. It is important to know when this time is reached in a 
collision and with which parameters. Then this will enable us to conclude
about the material parameters and the equilibration dynamics. The exact model
provides us with a simple and straightforward tool to give a precise
estimate about the time moment when the rotation equilibrated and the
parameters of the matter at that moment.  

The exact model can be used as a tool, when the rotations can be detected at 
freeze out. Then it provided an estimate of the rate of decrease
of angular speed and rotational energy due to the
expansion in an explosively expanding system.
This indicates
that the effects of rotation can be observable in case of
rapid freeze out and hadronization, and the rate if conversion
from rotational energy to expansion can be studied in detail
depending on the parameters of the model.

At the same time these studies also show that the initial
rotation is also influencing the rate of decrease of
rotation. Here especially the enhancement of initial
rotation due to the  Kelvin
Helmholtz Instability is essential, although this is a
special (3+1)D instability, which in itself
cannot be included in the exact rotational model.
Still  the presence of the KHI is essential to generate
 the rotation, and thus the observation
of the rotation is strongly connected to the evolving
turbulent instability in low viscosity Quark-gluon plasma.

Due to the difference of the time evolution between the
numerical and the exactly solvable hydrodynamical models at early times 
one expects that the predictions of the two models in the sector of 
penetrating probes (dilepton spectrum, direct photon spectra, elliptic 
flow, HBT) will be different and these differences later on can be used 
to explore the mechanism of equilibration.

\section*{Acknowledgements}

Enlightening discussions with Weibing Deng are gratefully acknowledged.
This research was partially supported by the Hungarian OTKA grant,
NK101438.


\end{document}